\documentclass[english,iop,preprint]{emulateapj}
\usepackage[T1]{fontenc}
\usepackage{graphicx}

\makeatletter
\usepackage{apjfonts}
\usepackage[hyperfootnotes=false,colorlinks=true,citecolor=blue]{hyperref}

\usepackage{apjfonts}
\usepackage[hyperfootnotes=false]{hyperref}
\hypersetup{colorlinks=true,citecolor=blue}
\shorttitle{The FUV Continuum Background shortward of Ly$\alpha$}
\shortauthors{SEON}

\makeatother

\usepackage{babel}
\begin{document}

\title{Lognormal intensity distribution of the FUV continuum background
shortward of Ly$\alpha$}

\author{Kwang-Il Seon\altaffilmark{1,2}}

\altaffiltext{1}{Korea Astronomy and Space Science Institute, Daejeon 305-348, Republic of Korea; kiseon@kasi.re.kr}
\altaffiltext{2}{Astronomy and Space Science Major, University of Science and Technology, Daejeon 305-350, Republic of Korea} 
\begin{abstract}
The diffuse far-ultraviolet (FUV) continuum radiation ``longward''
of Ly$\alpha$ (1216\AA) is well known to correlate with the dust
emission at 100 $\mu$m. However, it has been claimed that the FUV
continuum background ``shortward'' of Ly$\alpha$ shows very weak
or no correlation with the 100 $\mu$m emission. In this paper, the
observational data of the diffuse FUV radiation by the \textit{Far
Ultraviolet Spectroscopic Explorer} is reexamined in order to investigate
the correlation between the diffuse FUV radiation shortward of Ly$\alpha$
and the 100 $\mu$m emission. Large fluctuations were confirmed in
the linear-linear correlation plots, but good correlations were found
in the log-log plots. The large fluctuations in the linear-linear
plots, and thus poor correlations, between the FUV and 100 $\mu$m
intensities were attributed to the lognormal property of the FUV intensity
distribution. The standard deviation of the intensity distribution
of the FUV radiation shortward of Ly$\alpha$ was found to be $\sigma_{\log I}=0.16-0.25$.
The result is consistent with that obtained not only for the FUV radiation
longward of 1216\AA, but also with the dust column density measurements
of various molecular clouds. This implies that most of the diffuse
FUV radiation shortward of Ly$\alpha$ is dust-scattered light in
the turbulent interstellar medium. The diffuse FUV data obtained from
the \emph{Voyager} missions was also investigated. However, much wider
random fluctuations were found compared with the \emph{FUSE} data,
which is most likely due to the systematic difficulties in data reduction
of the \emph{Voyager} data.
\end{abstract}


\section{Introduction}

The far-ultraviolet (FUV) continuum background in the wavelength longer
than Ly$\alpha$ (1216\AA; hereafter, FUV-L) has been investigated
extensively and found to correlate with the 100 $\mu$m emission and
neutral hydrogen (\ion{H}{1}) column density, and thus to be mostly
starlight scattered by interstellar dust \citep{PMB1980,Bowyer91,Witt1997,SCH2001,Murthy2010,Seon2011a,Seon2011b}.
Since gas and dust are well mixed in the interstellar medium (ISM)
\citep[e.g., ][]{Boulanger,Cox89} and the intensity of the scattered
light is roughly proportional to the dust column density in the optically
thin limit, the diffuse scattered light should exhibit a signature
of the ISM density structure. The probability distribution functions
(PDFs) of the three-dimensional densities and column densities of
the turbulent ISM are well known to be close to lognormal not only
through numerical simulations \citep{Vazquez94,Nordlund99,Klessen2000,Ostriker01,Wada2001,Burkhart2012}
but also through observations \citep{Lada94,Berkhuijsen2008,Hill08,Seon2009,Padoan97,Lombardi2006,Lombardi2008,Kainulainen2009,Froebrich2010,Schneider2013}.
Recently, \citet{Seon2011a} found evidence that the PDF of the FUV-L
continuum background intensity is lognormal, and this is attributable
to the turbulent nature of the ISM.

Observations of the diffuse continuum background at the FUV wavelength
band shortward of Ly$\alpha$ (hereafter, FUV-S) have been relatively
scarce. The earliest observations of the FUV-S background were those
of \citet{Belyaev1971} from \emph{Venera 5} and \emph{6}. Further
photometric observations obtained the upper limits of the diffuse
FUV-S continuum background \citep{Henry1973,Paresce1976,Bixler1984,Opal1984}.
The most extensive observations of the FUV-S continuum background
came from the ultraviolet spectrographs (UVSs) aboard the two \emph{Voyager}
spacecraft \citep{Sandel79,Holberg86,Murthy91,Murthy99,Murthy2012}.
The \emph{Voyager} 1 and 2 UVSs observed the diffuse FUV background
from 1977 to 2001 and 1998, respectively. The entire data set of the
observations were published by \citet{Murthy2012}. \citet{Murthy99}
did not find a correlation between the diffuse FUV-S background and
the \ion{H}{1} column density. It was also claimed that the FUV-S
background correlates with the integrated stellar fluxes, which presumably
trace the local radiation field in each line of sight. However, several
difficulties in the \emph{Voyager} data set and/or the analyses have
been identified (\citealt{Edelstein2000}; see Section 3.2 for details),
and these may prevent proper interpretation of the \emph{Voyager}
results. \citet{Murthy04} used serendipitous observations from the
\emph{Far Ultraviolet Spectroscopic Explorer (FUSE)} to probe the
diffuse FUV-S background. Only a very weak correlation was found between
the FUV-S flux and the dust 100 $\mu$m emission; however, it had
large fluctuations and it was concluded that the FUV-S sky is very
patchy with both dark and bright regions. These results were attributed
to the differences in the local radiation field.

Meanwhile, it should be noted that even the FUV-L continuum and 100
$\mu$m intensities show large fluctuations in the linear-linear correlation
plots due to the lognormality of their intensity PDFs \citep{Seon2011a}.
\citet{Seon2011a} found that the correlation of the FUV-L continuum
background with the local stellar radiation field was less significant
than with the dust thermal emission at 100 $\mu$m. Therefore, it
can be speculated that the weak or no correlation between the FUV-S
continuum and 100 $\mu$m intensities is related to the lognormal
properties of the FUV-S intensity PDF. This was a key motivation to
reexamine the FUV-S background data observed with the \textit{FUSE}
mission \citep{Murthy04} in order to determine whether the large
fluctuations of the FUV-S background data could be explained using
the lognormal density PDF of the ISM. The data obtained from the \emph{Voyager}
\citep{Murthy2012} missions was also reexamined for completeness.
In Section 2, the correlation properties between two lognormal random
variables are investigated. The diffuse FUV-S continuum data are analyzed
in Section 3. Finally, concluding remarks are given in Section 4.

\section{Correlation between Two Lognormal Variables}

As will be seen in Section 3, when the correlation between the FUV-S
and 100 $\mu$m intensities is examined in a linear-linear plot, a
higher variance of the FUV-S intensity is found as the 100 $\mu$m
intensity increases, and vice versa. This tendency can be understood
if the PDFs of the two random variables, which are linearly dependent
on each other, are lognormal. A lognormal PDF for variable $I$ is
defined as follows:

\begin{equation}
P(\ln I)d\ln I=\frac{1}{\sqrt{2\pi}\sigma_{\ln I}}\exp\left[-\frac{\left(\ln I-\left\langle \ln I\right\rangle \right)^{2}}{2\sigma_{\ln I}^{2}}\right]d\ln I.\label{eq:lognormal_pdf}
\end{equation}
Here, the mean value $\left\langle \ln I\right\rangle $ of the logarithmic
quantity $\ln I$ is related to the linear mean value $\left\langle I\right\rangle $
by $\left\langle I\right\rangle =\left\langle \ln I\right\rangle e^{(1/2)\sigma_{\ln I}^{2}}$.
The variance ($\sigma_{\ln I}^{2}$) of $\ln I$ is related to the
variance ($\sigma_{I}^{2}$) or ``relative'' variance ($\sigma_{I/\left\langle I\right\rangle }^{2}=\sigma_{I}^{2}/\left\langle I\right\rangle ^{2}$)
of $I$ by $\sigma_{I/\left\langle I\right\rangle }^{2}=\exp\left(\sigma_{\ln I}^{2}\right)-1$
or $\sigma_{\ln I}^{2}=\ln(\sigma_{I/\left\langle I\right\rangle }^{2}+1)$.
Therefore, a constant variance of $\ln I$ ($\sigma_{\ln I}^{2}=$
constant) yields a constant relative variance ($\sigma_{I/\left\langle I\right\rangle }^{2}=$
constant) of $I$, and thus a linearly increasing standard deviation
of $I$ with mean value $\left\langle I\right\rangle $, i.e. $\sigma_{I}\propto\left\langle I\right\rangle $.

\begin{figure*}[t]
\begin{centering}
\includegraphics[clip,scale=0.9]{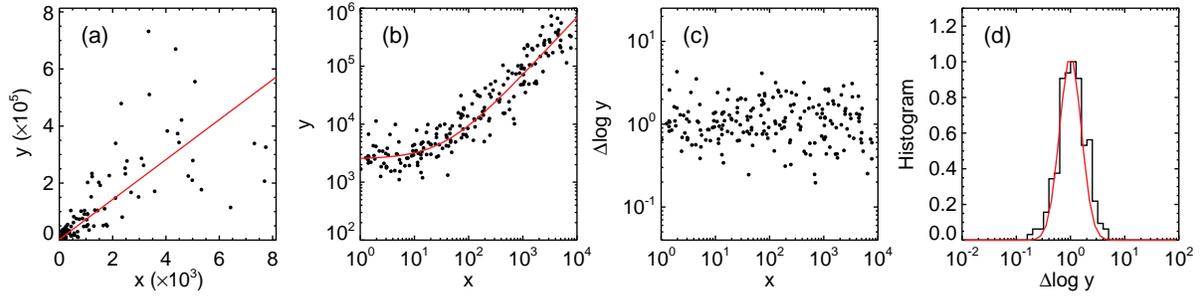}
\par\end{centering}

\caption{\label{simplot}Correlation plots between two lognormal variables,
which are linearly dependent on each other, in (a) linear-linear and
(b) log-log scales. Two lognormal variables ($x$ and $y$) were assumed
to be correlated according to the linear relationship $\left\langle y\right\rangle =70\left\langle x\right\rangle +2500$,
which is denoted by the red curves in (a) and (b). The standard deviations
of $\log x$ and $\log y$ were assumed to be $\sigma_{\log x}=\sigma_{\log y}=0.2$.
(c) The residual $\Delta\log y\equiv\log y-\left\langle \log y\right\rangle $
versus $\log x$, and (d) the histogram of $\Delta\log y$. The red
curve in (d) shows a lognormal function with $\sigma_{\log y}=0.2$.}
\end{figure*}

This property causes a rapidly increasing fluctuation with intensity
in a linear-linear correlation plot between two lognormal variables.
If two random variables (i.e. $x$ and $y$) are linearly correlated
on average, i.e.
\begin{equation}
\left\langle y\right\rangle =a\left\langle x\right\rangle +b,\label{eq:linear}
\end{equation}
and their PDFs are lognormal, the correlation plot between them would
exhibit large variances. Figure \ref{simplot} demonstrates correlation
between two lognormal variables in which the two variables $x$ and
$y$ were assumed to be related by the linear equation $\left\langle y\right\rangle =70\left\langle x\right\rangle +2500$,
as denoted by the red curves. The mean value $\left\langle x\right\rangle $
varied from 1.5 to 6400 in steps of 0.018 dex, and the two quantities
($x$ and $y$) were randomly generated to have lognormal distributions
with standard deviations of $\sigma_{\log x}=\sigma_{\log y}=0.2$.
Figures \ref{simplot}(a) and \ref{simplot}(b) present the correlation
plots between $x$ and $y$, and between $\log x$ and $\log y$,
respectively. It is evident that the correlation plot between the
linear quantities shows no clear correlation, but rather large fluctuations
for a given range of $x$, while the plot between the logarithmic
quantities shows a clear correlation. 

Therefore, the correlation between two lognormal quantities should
be examined with logarithmic values. In the log-log plot, the linear
relationship between the two logarithmic quantities may then be obtained
by fitting the two quantities to the following equation:

\begin{equation}
\left\langle \log y\right\rangle =\log\left(a10^{\log x}+b\right),\label{eq:log}
\end{equation}
 which is equivalent to Equation (\ref{eq:linear}). The standard
deviation or variance of $\log y$ can also be derived by fitting
the PDF of $\Delta\log y\equiv\log y-\left\langle \log y\right\rangle $
to a Gaussian function. Figures \ref{simplot}(c) and \ref{simplot}(d)
present the $\Delta\log y$ versus $\log x$ and the PDF of $\Delta\log y$
plots, respectively. The red curve in Figure \ref{simplot}(d) is
the Gaussian PDF for $\Delta\log y$ with a standard deviation of
0.2, which corresponds to the standard deviation of $\log y$. In
this manner, it will be shown that the FUV-S continuum background
is well described using a lognormal PDF and the standard deviation
of the PDF is derived in the following section.

\section{Data Analysis}

\subsection{FUSE data}

The \emph{FUSE} mission is described in detail by \citet{Moos00}.
The present study used the \emph{FUSE} S405/505 program data that
was reduced by \citet{Murthy04}, instead of reprocessing the raw
data. The S405/505 program observed the blank sky regions near a number
of alignment stars. \citet{Murthy04} integrated the FUV-S fluxes
over six bands ($\lambda\lambda$987--1021, $\lambda\lambda$1035--1081,
$\lambda\lambda$1100--1134, $\lambda\lambda$1134--1180, $\lambda\lambda$1175--1142,
and $\lambda\lambda$1129--1195) that were selected to avoid airglow
lines, and the diffuse astronomical signal data was tabulated as presented
in their Table 2. The six bands are denoted as ``F1'' to ``F6'', and
the total intensity averaged over the six bands is ``TOT''. In the
present analysis, the Improved Reprocessing of the IRAS Survey (IRIS)
map is used for the dust thermal emission at 100 $\mu$m \citep{Miville-Deschenes2005}.
The 100 $\mu$m map from \citet{Schlegel98} was also used and no
significant differences were found when compared with the results
obtained with the IRIS map. The main difference between the IRIS map
and the \citet{Schlegel98} map is that the IRIS map was obtained
by taking into account the variation of the detector gain with brightness
at scales smaller than 1$^{\circ}$ while a constant gain factor was
assumed in the \citet{Schlegel98} map.

\begin{figure*}[t]
\begin{centering}
\includegraphics[clip,scale=0.82]{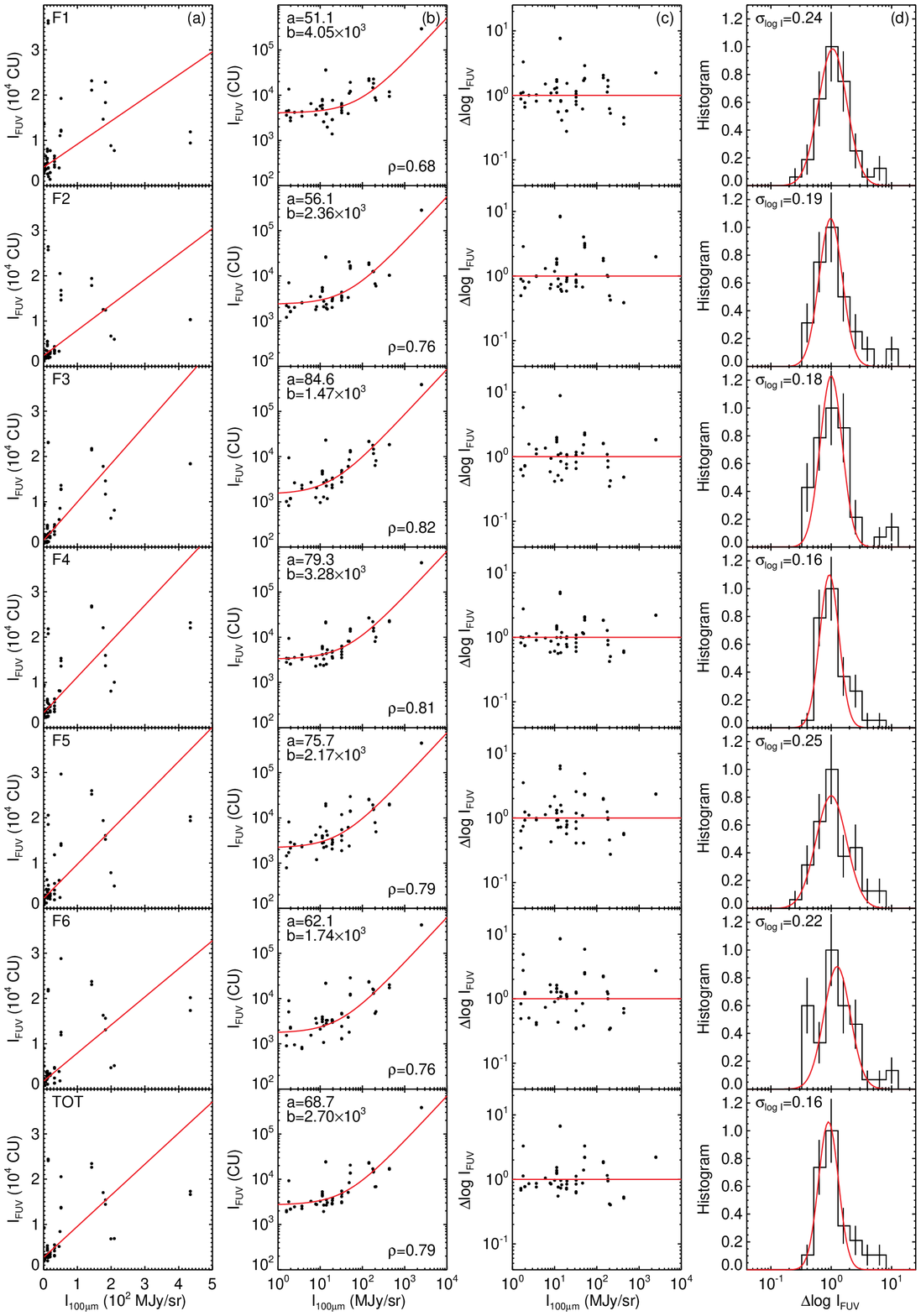}
\par\end{centering}

\caption{\label{correlation_plot}Correlation plots between the diffuse FUV-S
intensity obtained using \emph{FUSE} and the 100 $\mu$m intensity:
(a) $I_{{\rm FUV}}$ versus $I_{{\rm IR}}$, (b) $\log I_{{\rm FUV}}$
versus $\log I_{{\rm IR}}$, (c) $\Delta\log I_{{\rm FUV}}\equiv\log I_{{\rm FUV}}-\left\langle \log I_{{\rm FUV}}\right\rangle $
versus $\log I_{{\rm IR}}$, and (d) the probability distribution
function of $\Delta\log I_{{\rm FUV}}$. The red curves in (a) and
(b) are linear regression lines. The best-fit parameters $a$ and
$b$, and the Pearson's correlation coefficients are also shown in
(b). The best-fit lognormal function of $\Delta\log I_{{\rm FUV}}$
is shown in red in (d). \textbf{Here, CU represents the continuum
unit (photons cm$^{-2}$ s$^{-1}$ Å$^{-1}$ sr$^{-1}$).}}
\end{figure*}

First, the FUV-S intensities in the six bands are plotted against
the 100 $\mu$m intensity in linear-linear plots in the first column
of Figure \ref{correlation_plot}. As noted by \citet{Murthy04},
the linear-linear plots show very weak correlations and large fluctuations
in the FUV-S intensity for a given range of the 100 $\mu$m intensity.
The correlation relation between the $\log I_{{\rm FUV}}$ and $\log I_{{\rm IR}}$
is shown in the second column of Figure \ref{correlation_plot}. It
is clear that the large fluctuations seen in the linear-linear plots
have largely disappeared, and the correlation between the FUV-S and
100 $\mu$m intensities is more distinct. The correlation between
two quantities, e.g. the FUV-S intensity ($I_{{\rm FUV}}$) and the
100 $\mu$m intensity ($I_{{\rm IR}}$), is usually investigated using
a linear equation between the two linear intensities, i.e. $\left\langle I_{{\rm FUV}}\right\rangle =a\left\langle I_{{\rm IR}}\right\rangle +b$.
However, in this study, the correlation relations were fitted with
Equation (\ref{eq:log}) and the resulting best-fit parameters are
presented in the second column of Figure \ref{correlation_plot}.
The best-fit curves are also shown in red in the first and second
columns. In the second column, the Pearson correlation coefficient
($\rho$), which is defined as the covariance of the two variables
divided by the product of their standard deviations, is also denoted.
The coefficient is normally sensitive to outliers and thus the strong
correlations in Figure \ref{correlation_plot} could be suspected
to be the results of the highest intensity data ($I_{100\mu{\rm m}}\sim2.5\times10^{3}$
MJy sr$^{-1}$). Ignoring the highest intensity data lowered the correlation
coefficients, but no more than $\sim0.1$. Therefore, the strong correlations
between the FUV-S and 100 $\mu$m intensities are not due to the highest
intensity data. The residuals $\Delta\log I_{{\rm FUV}}\equiv\log I_{{\rm FUV}}-\left\langle \log I_{{\rm FUV}}\right\rangle $,
after subtracting the linear dependency of the FUV-S intensity on
the 100 $\mu$m intensity, versus $\log I_{{\rm IR}}$ are shown in
the third column in Figure \ref{correlation_plot}. It is noted that
the residuals $\Delta\log I_{{\rm FUV}}$ are more or less independent
of $\log I_{{\rm IR}}$. However, if the residuals were drawn in linear
scale, they would exhibit extended asymmetric tails to high intensity
values. This property is a characteristic of the lognormal PDF.

In order to confirm that the fluctuations from the average linear
dependence are represented by lognormal distributions, the histograms
of the residuals $\Delta\log I_{{\rm FUV}}$ were fitted with a Gaussian
function. In the fit, the Poisson error ($\sigma_{i}=\sqrt{N_{i}}$
in the bin $i$) was assumed, where $N_{i}$ is the number of data
points in the bin. The results are shown in the last column of Figure
\ref{correlation_plot}. It is clear that the PDFs are well represented
by the lognormal distributions, except some of the highest intensity
bins. The standard deviations of the logarithmic FUV-S intensities
$\sigma_{\log I}=0.16-0.25$ or $\sigma_{\ln I}=(\ln10)\sigma_{\log I}=0.37-0.58$
were obtained from the residual distributions. The standard deviations
in each wavelength band are also shown in the figure. Then, the relative
standard deviations or contrasts of the FUV-S intensity are given
by $\sigma_{I/\left\langle I\right\rangle }=0.38-0.63$.

\begin{figure*}[t]
\begin{centering}
\includegraphics[clip,scale=0.9]{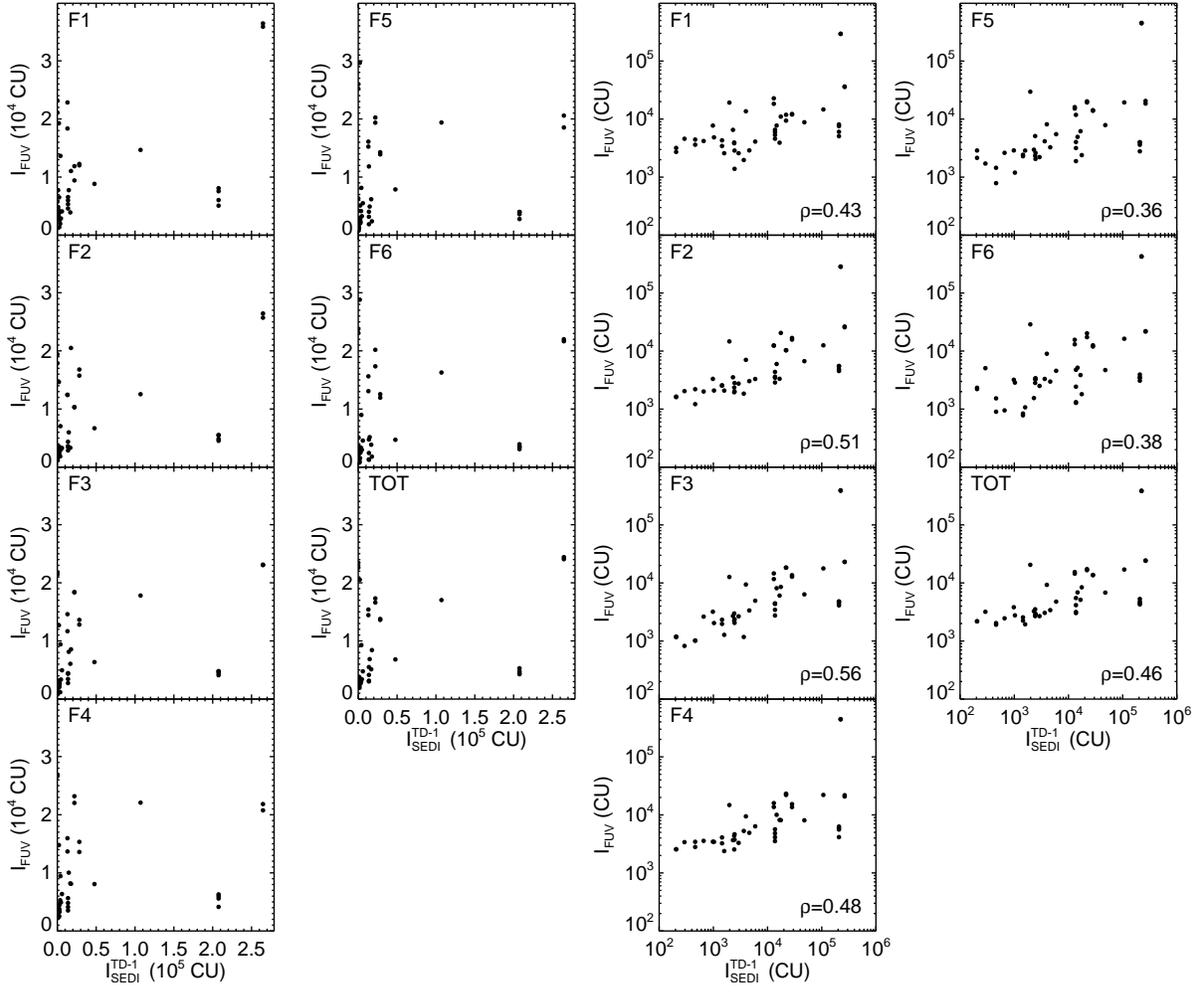}
\par\end{centering}

\caption{\label{correlation_td1}Correlation plots between the diffuse FUV-S
intensity obtained using \emph{FUSE} and the TD-1 SEDI. The linear-linear
plots are shown in the first and second columns; the log-log plots
are shown in the third and fourth columns. The Pearson's correlation
coefficients are also shown in the third and fourth columns.}
\end{figure*}

\citet{Murthy99} claimed that there is a strong correlation of the
FUV-S intensity obtained using the $Voyager$ UVSs with the stellar
flux from the OB stars, which presumably traces the local FUV radiation
in each line of sight, while there is no correlation with tracers
of the ISM. Hence, the correlation of the FUV-S intensity with the
stellar flux at $\sim$1565 \AA\ is examined using the TD-1 stellar
catalog \citep{Thompson1978}, as done by \citet{Murthy99}. The TD-1
catalog presents the absolute UV fluxes in four wavelength bands (centered
at 1565, 1965, 2365, and 2740 \AA, each being 310\textendash{}330
\AA\ wide). The stellar fluxes within 2$^{\circ}$ of the observed
position were integrated in order to calculate the \textquotedblleft{}stellar
equivalent diffuse intensity (SEDI)\textquotedblright{} \citep{Hurwitz1991,Seon2011a}.
Figure \ref{correlation_td1} presents the correlation plot between
the FUV-S intensity and the TD-1 SEDI. The Pearson correlation coefficients
are also denoted in the third and fourth columns. It is clear that
there is no apparent correlation in the linear-linear plots, while
the log-log plots reveal some correlations. It is also noted that
the correlations with the stellar flux are weaker than those with
the 100 $\mu$m intensity. A weaker correlation in the FUV-L intensity
with the local stellar flux than that with the ISM tracers was also
found \citep{Seon2011a}. These results do not support the previous
claim that the diffuse FUV continuum background correlates strongly
with the local stellar radiation field.

\subsection{Voyager data}

\begin{figure*}[t]
\begin{centering}
\includegraphics[clip,scale=0.9]{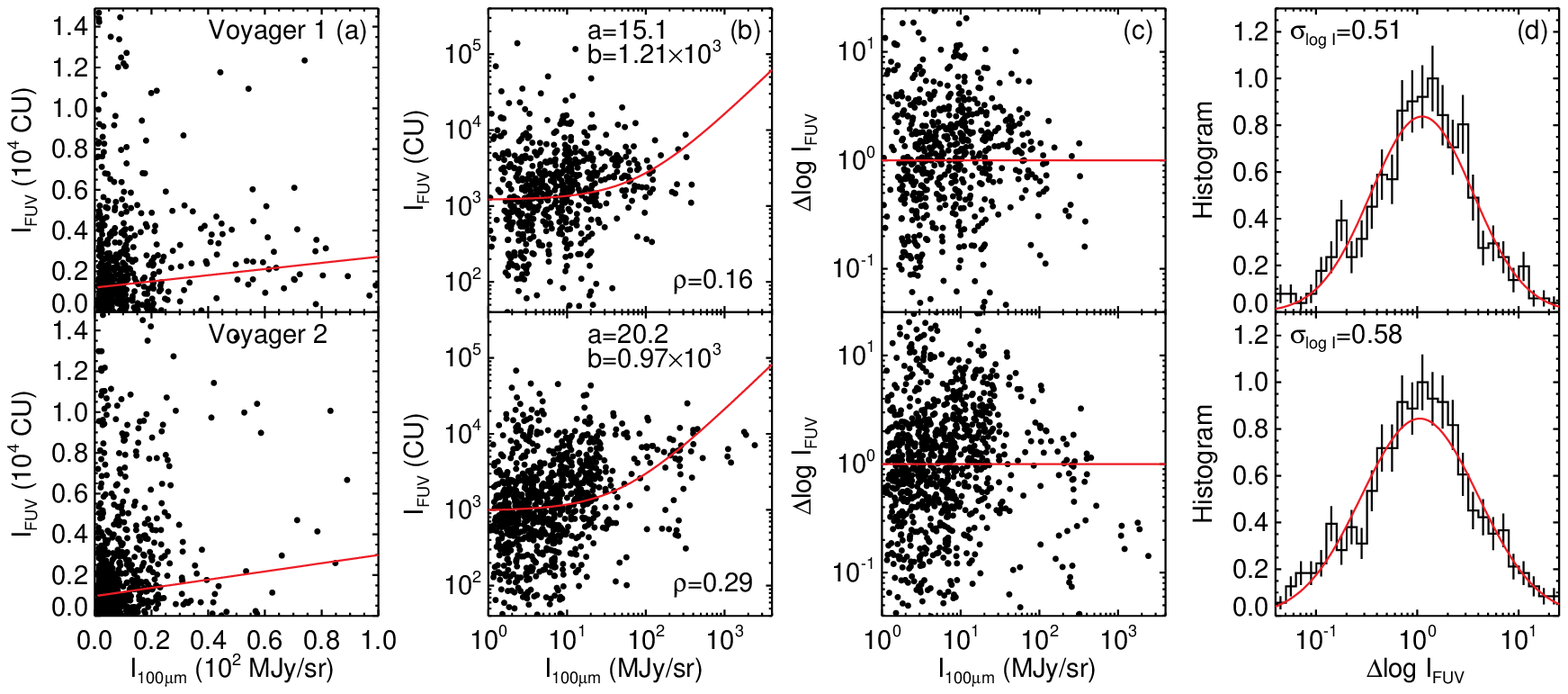}
\par\end{centering}

\caption{\label{voyager_corr}Correlation plots between the diffuse FUV-S intensity
obtained using \emph{Voyager} and the 100 $\mu$m intensity: (a) $I_{{\rm FUV}}$
versus $I_{{\rm IR}}$, (b) $\log I_{{\rm FUV}}$ versus $\log I_{{\rm IR}}$,
(c) $\Delta\log I_{{\rm FUV}}\equiv\log I_{{\rm FUV}}-\left\langle \log I_{{\rm FUV}}\right\rangle $
versus $\log I_{{\rm IR}}$, and (d) the probability distribution
function of $\Delta\log I_{{\rm FUV}}$. The red curves in (a) and
(b) are the linear regression lines. The best-fit parameters $a$
and $b$, and the Pearson's correlation coefficients are also shown
in (b). The best-fit lognormal function of $\Delta\log I_{{\rm FUV}}$
is shown in red in (d).}
\end{figure*}

The most extensive observations of the FUV-S continuum background
are provided by the two \emph{Voyager} UVSs. The Voyager UVSs provide
spectral data of the diffuse FUV radiation in the wavelength range
of $\sim540$--1700\AA. Therefore, the same analyses were performed
on the \emph{Voyager} FUV-S data in the wavelength range of 987--1200\AA\
(obtained from \citealt{Murthy2012}) as on the \emph{FUSE} data.
Figure \ref{voyager_corr} presents the linear-linear and log-log
correlation plots, $\Delta\log I_{{\rm FUV}}$ versus $I_{100\mu{\rm m}}$,
and the $\Delta\log I_{{\rm FUV}}$ histogram, as in Figure \ref{correlation_plot}
for the \emph{FUSE} data. The Pearson's correlation coefficients are
also shown with the log-log correlation plots. Unlike the \emph{FUSE}
data, the correlation between the \emph{Voyager} data and the 100
$\mu$m emission is very weak not only in the linear-linear plots,
but also in the log-log plots. The best-fit slopes ($a$) are less
steep than those found for the \emph{FUSE} data because the \emph{Voyager}
data were mostly obtained at sightlines with low 100 $\mu$m intensity.
However, the PDFs of the residuals from the best-fit linear curves
can still be represented using lognormal functions, but with much
broader widths than those for the \emph{FUSE} data. The best-fit curves
in Figure \ref{voyager_corr}(b) do not seem to follow the data at
first glance. Robustness of the results was assessed by varying the
initial parameter values for regression. Results were always the same
regardless of the adopted initial values. The symmetrical histograms
of the residuals in Figure \ref{voyager_corr}(d) also indicate the
robustness of the results.

\begin{figure*}[t]
\begin{centering}
\includegraphics[clip,scale=0.9]{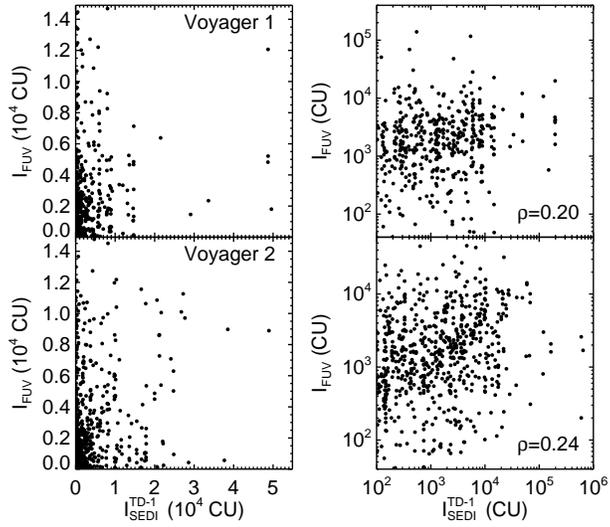}
\par\end{centering}

\caption{\label{voyager_td1}Correlation plots between the diffuse FUV-S intensity
obtained using \emph{Voyager} and the TD-1 SEDI. The linear-linear
plots are shown in the first column; the log-log plots are shown in
the second column. The Pearson's correlation coefficients are also
shown in the second column.}
\end{figure*}

Figure \ref{voyager_td1} shows the correlation between the FUV-S
intensity and the local radiation field. It is noted that the correlation
of the \emph{Voyager} 1 data with the local radiation field is slightly
better than that with the dust thermal emission, but not as strong
as \citet{Murthy99} claimed. However, the correlation of the \emph{Voyager}
2 data with the local radiation field is slightly worse than that
with the dust emission. Therefore, it is not plausible to discriminate
using the \emph{Voyager} data whether the correlation of the FUV-S
intensity with the 100 $\mu$m intensity is better or not than that
with the local radiation field.

The poor correlation between the \emph{Voyager} FUV-S data and the
100 $\mu$m emission, and the large fluctuation of the FUV-S intensity
even in the log-log plots may be attributable to unknown difficulties
in the data reduction of the \emph{Voyager} data \citep{Edelstein2000}.
The measurement of the diffuse flux with the \emph{Voyager} UVS data
requires complex corrections for noise components that are much larger
than the astronomical signal. There are three components in a typical
\emph{Voyager} spectrum: instrumental dark noises from the spacecraft's
radioisotope thermoelectric generator, four interplanetary emission
lines and their associated broad scattering wings, and cosmic signals
(which originate beyond the solar system). The most prominent emission
line feature in the observed spectrum is the heliospheric Ly$\alpha$
at 1216\AA, which is reflected by the calibration plate into the
spectrometer. The Ly$\alpha$ feature is so intense that its scattering
wings extend over almost the entire spectrum. Although \citet{Murthy91,Murthy99,Murthy2012}
claim that the noise components are relatively well determined, \citet{Edelstein2000}
note that there is still a significant danger of systematic effects,
which have not been fully appreciated in the data reduction procedures
of the \emph{Voyager} data, which dominate the random effects.

In order to confirm the systematic effects in the \emph{Voyager} data
reduction, the \emph{Voyager} FUV-L data obtained in the wavelength
range of $\sim1300$--1700\AA\ was also analyzed. Because the FUV-L
continuum intensity is well known to correlate with the 100 $\mu$m
intensity, the \emph{Voyager} FUV-L data should exhibit a good correlation
with the 100 $\mu$m intensity if there is no serious problem in the
data reduction of the \emph{Voyager} data. The Pearson's correlation
coefficients between the FUV-L intensity and the 100$\mu$m intensity
were $-0.06$ and 0.12 for the \emph{Voyager} 1 and 2 data, respectively.
These values are worse than that for the FUV-S data. Therefore, it
is highly likely that the systematic effects yielded large random
noises in the diffuse FUV-S fluxes of the \emph{Voyager} and hampered
the correlation analysis between the FUV-S background and the ISM
tracers. Thus, the results obtained using the \emph{Voyager} UVS spectrometers
are not discussed further in this paper.

\section{Concluding Remarks}

It was found that the intensity distribution of the FUV-S continuum
background obtained using the \emph{FUSE} mission is well represented
by a lognormal distribution. The linear-linear correlation plots of
the dust emission at 100 $\mu$m with the FUV-S radiation exhibited
large fluctuations, which were ascribed to the lognormal PDF properties
of the FUV-S background. The lognormal PDF exhibits a high value tail
and its standard deviation is proportional to the mean value. The
proportionality of the standard deviation to the mean value is the
principal reason for the large fluctuations in the linear-linear correlation
plots between the two independent lognormal variables.

As described in \citet{Seon2011a}, such large fluctuations, which
are evidence of the lognormal intensity distribution, have also been
noted in many previous studies of the diffuse FUV-L background radiation.
\citet{Joubert83} found that the distributions of the data points
in the linear-linear plots between the FUV-L intensity and \ion{H}{1}
column density were not symmetrical with respect to the linear correlation
lines. There were high intensity tails in the distributions of the
FUV-L intensities at a given \ion{H}{1} column density. These high
intensity tails in the FUV intensity histograms are also shown in
Figure 2 of \citet{PMB1980}. In the previous studies, the points
with the FUV-L intensity excesses were removed in the correlation
studies between the FUV-L intensity and ISM tracers \citep{Joubert83,Perault91}.
\citet{Joubert83} attributed this property to excess FUV radiation
in certain regions of the sky, perhaps due to two-photon emission
by a warm ionized medium (WIM). However, \citet{Deharveng82}, \citet{Reynolds92}
and \citet{Seon2011a} demonstrated that the emission from the WIM
is unlikely to contribute significantly to the diffuse FUV-L background.
\citet{SCH2001} also noted a large fluctuation in the FUV-L background
intensity and attributed the observed fluctuation to a natural consequence
of the multiple dust-scattering by several clouds along the line of
sight. \citet{Seon2011a} attributed these properties to the lognormal
nature of the FUV-L intensity PDF. The present study also revealed
the lognormal properties of the FUV-S intensity PDF.

Variations of the local radiation field would also affect the large
variances of the intensity PDF in the diffuse FUV background. This
might be particularly true in the present data that was obtained from
a wide range of the Galactic directions. However, it was found that
the correlation of the FUV-S intensity with dust is better than that
with the local radiation field. A similar result was also found for
the FUV-L background in \citet{Seon2011a}. It is also noted that
even in a relatively small area, where the stellar radiation field
might be more or less uniform, the FUV-L intensity exhibited a larger
variation at a higher intensity (e.g. see Figure 3 of \citealt{Park_SJ2012}).
The stellar radiation field impinging on a location in space is attenuated
by dust between the surrounding stars and the location. Hence, the
local radiation field in a region is modulated by the dust density
structure between the surrounding stars and the region. This modulated
incident radiation field is scattered by dust grains in the area,
and the scattered light is modulated again by the dust density structure
in the area. Then, the resulting fluctuation of the dust-scattered
light observed from the area is a convolution of the dust density
structure in the area and the modulation of the incident local radiation
field. Therefore, the observed FUV background should exhibit large
variances that are caused by the statistical properties of the ISM
density structure.

Numerical simulations have shown that the PDFs of the ISM density
and column density are close to lognormal and are caused by supersonic
compressible turbulence \citep{Vazquez94,Nordlund99,Klessen2000,Ostriker01,Wada2001,Burkhart2012}.
Observations of the various ISM phases have also revealed the lognormal
properties of the ISM density PDFs when the density structures are
mainly governed by turbulence. The PDFs of the \ion{H}{1} column
density of the Large Margellanic Cloud and the Milky Way were found
to be lognormal \citep{Lada94,Berkhuijsen2008}. The H$\alpha$ emission
measures of the Milky Way and M 51 were also well represented by lognormal
distributions \citep{Hill08,Berkhuijsen2008,Seon2009}. \citet{Padoan97}
found that the variation of the stellar extinction is consistent with
the lognormal distribution of the dust density. The PDFs of the dust
column density for many molecular clouds were well fitted using the
lognormal functions in low density (or turbulence dominant) regimes
\citep{Lombardi2006,Lombardi2008,Kainulainen2009,Froebrich2010,Schneider2013}.
However, it should be noted that the PDF shape becomes more complicated
when other physical processes (i.e., gravity, magnetic fields, feedback
effects like compression) apart from turbulence become important.
The PDFs of star-forming clouds show power-law tails, most likely
caused by gravity \citep[e.g.,][]{Kainulainen2009,Schneider2013}.

\begin{figure*}[t]
\begin{centering}
\includegraphics[clip,scale=0.72]{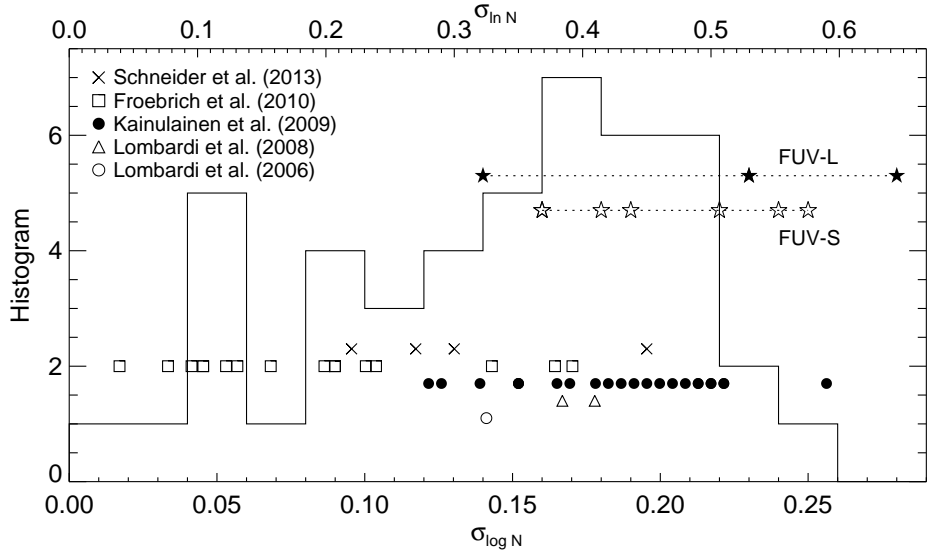}
\par\end{centering}

\caption{\label{sigma}Distribution of the standard deviations ($\sigma_{\log N}$
or $\sigma_{\ln N}$) of the dust column density ($N$) derived from
various molecular clouds. The standard deviations from individual
molecular clouds are also shown; their $y$-values were chosen arbitrarily
for clarity. The standard deviations derived from the FUV-S and FUV-L
continuum background are also shown.}
\end{figure*}

It is known that the standard deviation of normalized density ($\sigma_{\rho/\left\langle \rho\right\rangle })$
is proportional to the sonic Mach number \citep{Nordlund99,Ostriker01,Federrath2008,Federrath2009,Federrath2010}.
The variance of the logarithmic column density ($\sigma_{\ln N}^{2}$)
is approximately proportional to that of density ($\sigma_{\ln\rho}^{2}$)
\citep{Burkhart2012,Seon2012}. The proportional constants depend
on the turbulence forcing type and the magnetic field strength. Because
the intensity of the scattered light is in general proportional to
the dust column density, the standard deviation of the scattered light
provides some information on the turbulent properties of the ISM.
In the present study, the standard deviation of the FUV-S intensity
PDF was found to be in the range of $\sigma_{\log I}\sim0.16-0.25$
or, equivalently, $\sigma_{\ln I}=(\ln10)\sigma_{\log I}=0.37-0.58$
or $\sigma_{I/\left\langle I\right\rangle }=0.38-0.63$. \citet{Seon2011a}
found that $\sigma_{\log I}\sim0.14-0.28$ and $\sigma_{I/\left\langle I\right\rangle }=0.34-0.72$
for the FUV-L background, which is consistent with the results of
the present study for the FUV-S background. Figure \ref{sigma} presents
the frequency histogram of the standard deviation $\sigma_{\log N}$
(or $\sigma_{\ln N}$) of the dust column density obtained from various
molecular clouds. In the figure, the standard deviation values ($\sigma_{\log N}$)
are also shown and their $y$ values were arbitrarily chosen for clarity.
The standard deviations obtained from the present study for the FUV-S
background and from \citet{Seon2011a} for the FUV-L background are
also shown for comparison. The most probable value of $\sigma_{\log N}$
obtained from molecular clouds is $\approx0.14-0.22$, which is consistent
with the present results for the FUV-S and FUV-L backgrounds. The
similarity between the dispersions of the PDFs strongly suggests that
the fluctuation of the diffuse FUV background is primarily caused
by the turbulence property of the ISM. The broadening effects due
to the differences in the local stellar radiation field are relatively
small.

More detailed statistical properties of the dust density distribution
should be investigated through a radiative transfer modeling of the
starlight scattered by dust grains in the turbulent ISM. The turbulent
ISM could be simulated by the fractional Brownian algorithm, as described
in \citet{Elmegreen02} and \citet{Seon2012}. Therefore, the statistical
properties in dust-scattered light will be investigated using the
fractional Brownian motion algorithm in future work.

The primary difference between the diffuse FUV-L and FUV-S background
radiations is the spectral types of stars that are responsible for
the scattered light in each wavelength band. Earlier type stars radiate
stronger FUV-S radiation and the scale height of the earlier type
stars is lower than that of later type stars. Another difference is
the FUV-S radiation being more easily extinguished by dust grains
than the FUV-L radiation. However, the statistical properties of the
dust density are irrelevant to the spectral types of stars. Radiative
transfer models with realistic dust and stellar distributions in three-dimensional
space may be required in order to better understand the diffuse FUV
continuum radiation.

\clearpage{}

\begin{thebibliography}{References}
\bibitem[Belyaev et al.(1971)]{Belyaev1971}Belyaev, V. P., Kurt,
V. G., Melioranskii, A. S., et al. 1971, Cosmic Research, 8, 677

\bibitem[Berkhuijsen \& Fletcher(2008)]{Berkhuijsen2008}Berkhuijsen,
E. M., \& Fletcher, A. 2008, \mnras, 390, L19

\bibitem[Bixler et al.(1984)]{Bixler1984}Bixler, J., Bowyer, S.,
\& Grewing, M. 1984, \aap, 141, 422

\bibitem[Boulanger \& P{\'e}rault(1988)]{Boulanger}Boulanger, F.,
\& P{\'e}rault, M. 1988, \apj, 330, 964

\bibitem[Bowyer(1991)]{Bowyer91}Bowyer, S. 1991, \araa, 29, 59

\bibitem[Burkhart \& Lazarian(2012)]{Burkhart2012}Burkhart, B., \&
Lazarian, A. 2012, \apjl, 755, L19

\bibitem[Cox \& Mezger(1989)]{Cox89}Cox, P., \& Mezger, P. G. 1989,
A\&AR, 1, 49

\bibitem[Deharveng et al.(1982)]{Deharveng82}Deharveng, J. M., Joubert,
M., \& Barge, P. 1982, \aap, 109, 179

\bibitem[Edelstein et al.(2000)]{Edelstein2000}Edelstein, J., Bowyer,
S., \& Lampton, M. 2000, \apj, 539, 187

\bibitem[Elmegreen(2002)]{Elmegreen02}Elmegreen, B. G. 2002, \apj,
564, 773

\bibitem[Federrath et al.(2008)]{Federrath2008}Federrath, C., Klessen,
R. S., \& Schmidt, W. 2008, \apjl, 688, L79

\bibitem[Federrath et al.(2009)]{Federrath2009}Federrath, C., Klessen,
R. S., \& Schmidt, W. 2009, \apj, 692, 364

\bibitem[Federrath et al.(2010)]{Federrath2010}Federrath, C., Roman-Duval,
J., Klessen, R. S., Schmidt, W., \& Mac Low, M.-M. 2010, \aap, 512,
A81

\bibitem[Froebrich \& Rowles(2010)]{Froebrich2010}Froebrich, D.,
\& Rowles, J. 2010, \mnras, 406, 1350

\bibitem[Henry(1973)]{Henry1973}Henry, R. C. 1973, \apj, 179, 97

\bibitem[Hill et al.(2008)]{Hill08}Hill, A. S., Benjmin, R. A., Kowal,
G., et al. 2008, \apj, 686, 363

\bibitem[Holberg(1986)]{Holberg86}Holberg, J. B. 1986, \apj, 311,
969

\bibitem[Hurwitz et al.(1991)]{Hurwitz1991}Hurwitz, M., Bowyer, S.,
\& Martin, C. 1991, \apj, 372, 167

\bibitem[Joubert et al.(1983)]{Joubert83}Joubert, M., Masnou, J.
L., Lequeux, J., Deharveng, J. M., \& Cruvellier, P., 1983, \apj,
338, 677

\bibitem[Kainulainen et al.(2009)]{Kainulainen2009}Kainulainen, J.,
Beuther, H., Henning, T., \& Plume, R. 2009, \aap, 508, L35

\bibitem[Klessen(2000)]{Klessen2000}Klessen, R. S. 2000, \apj, 535,
869

\bibitem[Lada et al.(1994)]{Lada94}Lada, C. J., Lada, E. A., \& Clemens,
D. P., \& Bally, J. 1994, \apj, 429, 694

\bibitem[Lombardi et al.(2006)]{Lombardi2006}Lombardi, M., Alves,
J., \& Lada, C. J. 2006, \aap, 454, 781

\bibitem[Lombardi et al.(2008)]{Lombardi2008}Lombardi, M., Alves,
J., \& Lada, C. J. 2008, \aap, 489, 143

\bibitem[Miville-Desch{\^e}nes \& Lagache(2005)]{Miville-Deschenes2005}Miville-Desch{\^e}nes,
M.-A., \& Lagache, G. 2005, \apjs, 157, 302

\bibitem[Moos et al.(2000)]{Moos00}Moos, H. W., Cash, W. C., Cowie,
L. L., et al. 2000, \apjl, 583, L1

\bibitem[Murthy et al.(1999)]{Murthy99}Murthy, J., Hall, D., Earl,
M., Henry, R. C., \& Holberg, J. B. 1999, \apj, 522, 904

\bibitem[Murthy et al.(1991)]{Murthy91}Murthy, J., Henry, R. C.,
\& Holberg, J. B. 1991, \apj, 383, 198

\bibitem[Murthy et al.(2012)]{Murthy2012}Murthy, J., Henry, R. C.,
\& Holberg, J. B. 2012, \apjs, 199, 11

\bibitem[Murthy et al.(2010)]{Murthy2010}Murthy, J., Henry, R. C.,
\& Sujatha, N. V. 2010, \apj, 724, 1389

\bibitem[Murthy \& Sahnow(2004)]{Murthy04}Murthy, J., \& Sahnow,
D. J. 2004, \apj, 615, 315

\bibitem[Nordlund \& Padoan(1999)]{Nordlund99}Nordlund, \AA., \&
Padoan, J. 1999, in Interstellar Turbulence, ed. J. Franco \& A. Carraminana
(Cambridge: Cambridge University Press), 218

\bibitem[Opal \& Weller(1984)]{Opal1984}Opal, C. B., \& Weller, C.
S. 1984, \apj, 282, 445

\bibitem[Ostriker et al.(2001)]{Ostriker01}Ostriker, E. C., Stone,
J. M., \& Gammie, C. F. 2001, \apj, 546, 980

\bibitem[Padoan et al.(1997)]{Padoan97}Padoan, P., Jones, B. J. T.,
\& Nordlund, \AA. 1997, \apj, 474, 730

\bibitem[Paresce \& Bowyer(1976)]{Paresce1976}Paresce, F., \& Bowyer,
S. 1976, \apj, 207, 432

\bibitem[Paresce et al.(1980)]{PMB1980}Paresce, F., McKee, C. F.,
\& Bowyer, S. 1980, \apj, 240, 387

\bibitem[Park et al.(2012)]{Park_SJ2012}Park, S.-J., Min, K.-W.,
Seon, K.-I., et al. 2012, \apj, 754, 10

\bibitem[P{\'e}rault et al.(1991)]{Perault91}P\'erault, M., Lequeux,
J., Hanus, M., \& Joubert, M. 1991, \aap, 246, 243

\bibitem[Reynolds(1992)]{Reynolds92}Reynolds, R. J. 1992, \apjl,
392, L35

\bibitem[Sandel et al.(1979)]{Sandel79}Sandel, B. R., Shemansky,
D. E., \& Broadfoot, A. L. 1979, \apj, 227, 808

\bibitem[Schiminovich et al.(2001)]{SCH2001}Schiminovich, D., Friedman,
P. G., Martin, C., \& Morrissey, P. F. 2001, \apjl, 563, L161

\bibitem[Schlegel et al.(1998)]{Schlegel98}Schlegel, D. J., Finkbeiner,
D. P., \& Davis, M. 1998, \apj, 500, 525

\bibitem[Schneider et al.(2013)]{Schneider2013}Schneider, N., Andre,
P., K{\"o}nyves, V., et al. 2013, \apjl, 766, L17

\bibitem[Seon(2009)]{Seon2009}Seon, K.-I. 2009, \apj, 703, 1159

\bibitem[Seon(2012)]{Seon2012}Seon, K.-I. 2012, \apjl, 761, L17

\bibitem[Seon et al.(2011a)]{Seon2011a}Seon, K.-I., Edelstein, J.,
Korpela, E. J., et al. 2011a, \apjs, 196, 15

\bibitem[Seon et al.(2011b)]{Seon2011b}Seon, K.-I., Witt, A., Kim,
I.-J., et al. 2011b, \apj, 743, 188

\bibitem[Thompson et al.(1978)]{Thompson1978}Thompson, G. I., Nandy,
K., Jamar, C., et al. 1978, Catalogue of Stellar Ultraviolet Fluxes.
A Compilation of Absolute Stellar Fluxes Measured by the Sky Survey
Telescope (S2/68) aboard the ESRO Satellite TD-1 (London: Sci. Res.
Council)

\bibitem[V{\'a}zquez-Semadeni(1994)]{Vazquez94}V\'azquez-Semadeni,
E. 1994, \apj, 423, 681

\bibitem[Wada \& Norman(2001)]{Wada2001}Wada, K., \& Norman, C. A.
2001, \apj, 547, 172

\bibitem[Witt et al.(1997)]{Witt1997}Witt, A. N., Friedmann, B. C.,
\& Sasseen, T. P. 1997, 481, 809

\end{thebibliography}
\end{document}